# Probing surface noise with depth-calibrated spins in diamond

B. A. Myers[1], M. C. Dartiailh[1], K. Ohno[1], A. Das[1], D. D. Awschalom[1,2], and A. C. Bleszynski Jayich[1]

[1]Department of Physics, University of California, Santa Barbara, California 93106, USA
[2]Institute for Molecular Engineering, University of Chicago, Chicago, Illinois 60637, USA

Sensitive nanoscale magnetic resonance imaging (MRI) of target spins using nitrogen-vacancy (NV) centers in diamond will require a quantitative understanding of dominant noise at the surface. We probe this noise by applying dynamical decoupling to shallow NVs at calibrated depths. Results support a model of NV dephasing by a surface bath of electronic spins having a correlation rate of 200 kHz, much faster than that of the bulk N spin bath. Our method of combining nitrogen delta-doping growth and nanoscale depth imaging paves a way for studying spin noise present in diverse material surfaces.

The negatively charged nitrogen-vacancy (NV) center in diamond is a robust quantum sensor of magnetic fields [1-4]. Although an individual NV has the capability to detect small numbers of electronic [5-7] and nuclear spins external to diamond [8-10], its widespread application in spin imaging has been limited by the ability to form shallow NVs that retain spin coherence near the surface. Shallow spins with long coherence time, $T_2$, are important because quantum phase accumulation between two electronic spin states of the NV provides signal transduction, and hence the minimum detectable magnetic dipole moment scales as $\delta\mu \propto r^3/\sqrt{T_2}$, with $r$ the NV-target spin distance [3,4]. At odds with this figure of merit is strong evidence that the diamond crystal surface adversely affects $T_2$, reducing it from ~2 ms for bulk NVs [11,12] to less than 10 µs for few-nm deep NVs [6,13-16], but the origin of this decoherence is an outstanding question. We consider in this letter a model of surface spin induced decoherence, a theory which has emerged from experiments on other systems [20,21] where long coherence is a requirement, such as in superconducting circuits [17,18] and spin qubits in silicon [19]. We show that an electronic surface spin model is quantitatively supported for NVs in diamond. The key step we present is to link NV coherence with precise, independently measured NV depth data, as enabled by recent advancements in depth-controlled NV center creation and nanometer-scale magnetic imaging.

Recently, Ohno et al. demonstrated shallow coherent NVs using delta-doping of nitrogen during chemical vapor deposition (CVD) of single-crystal diamond (SCD) [16]. This crystal growth technique both permits nm-scale depth confinement and minimizes crystal damage incurred during nitrogen ion implantation [13,15,22], the conventional method of generating shallow NVs. The long $T_2$ of these doped NVs has enabled detection of a nanoscale volume of actively manipulated external protons [10]. The consistent NV quality in delta-doped SCD makes depth measurements a suitable probe of surface physics, not masked by effects of other process-induced crystal variations. Therefore we used this promising material in the reported work: we exploit depth-calibrated NVs to understand how the surface contributes to decoherence and provide a way to mitigate surface noise for enhanced external spin sensing. Using CPMG based dynamical decoupling (DD) for coherence analysis [23], we varied the number of CPMG pulses to deduce the noise spectral contributions from the surface and bulk environments as a function of depth. We show that CPMG with shorter inter-pulse spacing can progressively increase efficiency in decoupling from rapid magnetic fluctuations at the surface.

We prepared shallow NVs – all within 160 nm of the surface – in three depth-confined layers of isotopically pure $^{15}$N ($>98\%$) within an isotopically purified $^{12}$C (99.999%) CVD-grown film, shown schematically in Fig. 1(a). We grew the SCD epitaxially using plasma-enhanced CVD with the conditions and post-growth NV formation in ref. [16,24]. All experiments were performed within a single grown diamond film, thereby eliminating sample-to-sample surface variations. Nanometer scale changes in an NV's depth are critical to both its magnetic sensitivity and spatial resolution; thus we require an independent method to discriminate NVs' depths beyond the diffraction-limited resolution afforded by standard confocal microscopy [25]. NV-based detection of nuclear spins prepared on the surface can infer an absolute NV depth, though analysis requires an assumed spin magnetic field model and the measurement is time intensive and inaccessible for all but the highest quality NVs sufficiently close to the surface [9,10]. Here we employ a magnetic field gradient assisted optically detected electron spin resonance (ODESR) imaging technique that resolves NV depth differences with nanometer resolution [1,26] over a wide depth range of several 100 nm. Moreover, no assumed model is necessary to extract relative NV depths. Absolute depths are inferred by linking this technique with a model of NV coupling to surface spins.

We identify NVs and their depths by combining an inverted confocal microscope and an atomic force microscope (AFM) with a probe magnetized along the tip axis [24]. Within the film of nitrogen delta-doped layers (Fig. 1(a)) we differentiate doped $^{15}$NVs from bulk, naturally occurring $^{14}$NVs through confocal fluorescence (Fig. 1(b)) and ODESR spectroscopy of the $^{15}$N hyperfine sublevels [16,27]. All data presented in this paper are on $^{15}$NVs, which are referred to as NVs. To image NV depths, the magnetic AFM tip was scanned over the diamond surface at constant height, producing a bowl-shaped scanning ESR slice near the NVs (Fig 1(a)). This slice corresponds to the locus of points in space where the magnetic field along a specified NV axis is constant and brings the NV $|m_s=0\rangle \leftrightarrow |m_s=-1\rangle$ transition, $\nu_{NV}$ = 2.87 GHz $-\gamma_{NV}\left(B_{DC}+B_{tip}\right)$, into resonance with a microwave field $\nu_{RF}$, where $\gamma_{NV}$ is the NV gyromagnetic ratio and $B_{DC}$ and $B_{tip}$ and are the externally applied and tip magnetic fields. When an NV intersects the slice, its fluorescence decreases due to its spin-dependent coupling into a long-lived metastable state. In this way, a single NV images the resonant slice, as shown in Figs. 1(c) and 2(a), where the dark contours correspond to the $(x,y,z)$ tip positions for which $v_{NV} \approx v_{RF}$.

We obtained relative depth between any two NVs by registering their $(y,z)$ resonance slice images (Fig 2(a)) and extracting the vertical offset. The relative depth for a given NV was computed from its mean offset from every other NV, and the standard error of the mean for each NV depth ranged 1-2 nm [24]. Figure 2(b) is a plot of the Hahn echo coherence envelopes for NVs at four distinct depths, showing $T_2$ decreases with proximity to the surface. The coherence decay envelope of an NV depends on the nature of the environmental spin bath, described by its noise spectral density $\tilde{S}(\omega)$, and the measurement microwave pulse sequence, which applies a filter function to $\tilde{S}(\omega)$. To further isolate the surface-specific contribution we applied higher order DD, specifically CPMG-$N$, to reduce the inter-pulse spacing $\tau$ for a given total precession time $T$ and thereby decouple the NV from fluctuating fields at frequencies $f < 1/\tau$. Hence, comparing NV coherence subject to different pulse sequences reveals the bath dynamics; here Hahn echo and XY4 were used for their relevance to magnetometry applications [28]. The data for $T_{2,echo}$ and $T_{2,XY4}$ versus depth are plotted in Fig. 2(c) for 13 NVs; the zero depth mark is at an estimated absolute position discussed later. Both $T_{2,echo}$ and $T_{2,XY4}$ increased nearly monotonically with depth and are suppressed most drastically for depths $<25$ nm. Importantly, the coherence enhancement $T_{2,XY4}/T_{2,echo}$ declined from a value of $\sim 2.52$ deep in the film to as low as 1.2 for a shallow NV (lower panel of Fig. 2(c)). This reduced decoupling efficiency of NVs near the surface suggests a depth-dependent change in the nature of the dominant spin bath from that of a homogeneous bulk bath to a faster fluctuating configuration of surface spins whose effects are not decoupled at longer precession times. Figure 2(c) also shows that the longitudinal spin relaxation time $T_1$ decreased for shallow NVs over a similar depth scale as the $T_2$ decrease, though $T_1$ was generally an order of magnitude larger than $T_{2,\text{echo}}$. Therefore $T_1$ processes did not dominate spin decoherence.

To explain the degradation of NV coherence near the surface, we assumed a noise model of pure spin dephasing [24]. Based on the saturation of $T_2$ in Fig 2(c) and the isotopically pure $^{12}CH_4$ growth precursor [16], we expected that the dephasing of NVs deep in the film was dominated by interactions with a bulk-like spin bath of nitrogen P1 centers. Such magnetic noise is well described by a mean field theory with the Ornstein-Uhlenbeck process [23], which phenomenologically has a Lorentzian spectral density centered at zero frequency. It is a natural ansatz to take the total noise spectral density for an NV to be a two-Lorentzian sum with contributions from the bulk and surface

$$\tilde{S}_{\text{bulk}}(\omega) + \tilde{S}_{\text{surf}}(\omega) = \frac{b_{\text{bulk}}^2}{\pi} \frac{\tau_{\text{bulk}}}{1+\omega^2 \tau_{\text{bulk}}^2} + \frac{b_{\text{surf}}^2}{\pi} \frac{\tau_{\text{surf}}}{1+\omega^2 \tau_{\text{surf}}^2}, \tag{1}$$

where $b_{\text{bulk}}$, $b_{\text{surf}}$ are the NV-noise bath coupling frequencies and $\tau_{\text{bulk}}$, $\tau_{\text{surf}}$ are the baths' autocorrelation times. The dephasing theory predicts a reduced coherence $C_N$ after total NV precession time $T$:

$$C_N(T,b,\tau_c) = \exp\left[-\int_{-\infty}^{\infty} d\omega \tilde{S}(\omega) \mathcal{F}_N(T,\omega)\right], \quad (2)$$

where $\mathcal{F}_N$ is a filter function for the specific $N$-pulse CPMG measurement (see supplementary information) [24]. We simultaneously fit Hahn echo and XY4 coherence decay data to $C_1$ (Hahn) and $C_4$ (XY4) and extracted parameters $b_{bulk}$, $b_{surf}$, $\tau_{bulk}$, and $\tau_{surf}$. For deep NVs ($d > \sim 60$ nm) we found that $b_{surf}$, and thus $\tilde{S}_{surf}(\omega)$, is negligible, and we determined parameters $b_{bulk} \approx 13$ kHz and $\tau_{bulk} \approx 830.2$ μs [24]. The $T_{2,XY4}/T_{2,echo} \approx 2.52$ of these NVs is consistent with $N^\lambda$, where $\lambda = 2/3$ is expected for a "slow bath" of fluctuating nitrogen spins [23]. This theory predicts from the measured $b_{bulk}$ a nitrogen density of $\rho_{bulk} = 8.6 \times 10^{15}$ cm$^{-3}$ (48.5 ppb), also consistent with our mean $T_{2,echo} = 410$ μs [24] and Secondary Ion Mass Spectrometry data on nitrogen concentration in the delta-doped films [16].

For spins closer to the surface we fit the coherence envelopes to the full two-bath model of Eq. 2, fixing $b_{bulk}$ and $\tau_{bulk}$ to the values found for deep spins. We found a depth-dependent $b_{surf}$ ranging $3-170$ kHz and a depth-independent $\tau_{surf} = 5(3)$ μs, corresponding to a faster bath than in the bulk and explaining why $T_{2,XY4}/T_{2,echo}$, and hence $\lambda$, was significantly reduced with NV proximity to the surface. The lack of depth-dependence in $\tau_{surf}$ is consistent with $\tau_{surf}$ being internal to the bath. The depth dependence of $b_{surf}$ is well described by a 2D layer of surface $g = 2$ spins, and furthermore, the model yields an absolute NV depth. By integrating over a uniform surface distribution $\sigma_{surf}$ of fluctuating $S = 1/2$ dipoles, we find the total mean square field along the NV axis:

$$B_{rms}^2 = b_{surf}^2(d)/\gamma_{NV}^2 = \left(\frac{g\mu_0\mu_B}{4\pi}\right)^2 \frac{13\pi}{96} \frac{\sigma_{surf}}{(d-d_0)^4}, \qquad (3)$$

where $d$ is the relative NV depth (arbitrary zero) and $d_0$ is an offset to find absolute depth [24]. A fit of Eq. (3) to the $b_{surf}$ data points in Fig. 3(a) predicts absolute depths $(d-d_0)$ of the shallowest two NVs at 6 nm and 9 nm, consistent with the growth rate; henceforth $d$ denotes absolute depth. We find a surface spin density $\sigma_{surf} = 0.04(2)$ spins/nm$^2$, corresponding to a $r_0 \approx 2.8$ nm mean spin separation. The non-discrete surface spin model is justified because $r_0 < d$ for all NVs studied. The depth dependence $b_{surf}(d) \propto 1/d^2$ is in good agreement with the $b_{surf}$ data. Other functions $1/d^\alpha$ using $\alpha < 2$ (e.g., semi-3D spin volume) or $\alpha \sim 3$ (single spin) fit less well to the deepest $b_{surf}$ values, highlighting the necessity of measuring NVs at a broad range of depths [24]. Figure 3(b) joins the shallow and bulk noise models in a plot of integrated noise power $b^2 = b_{bulk}^2 + b_{surf}^2$, which can also be expressed in magnetic field units as $B^2 = b^2/\gamma_{NV}^2$. The sharp increase in $b^2$ reflects the decrease in spin coherence times at $d < 25$ nm in Fig. 2(c), and therefore 25 nm is approximately the depth at which rapidly fluctuating surface spins, rather than the slow P1 spin bath, begin to dominate NV decoherence.

To mitigate surface noise and enhance magnetic sensitivity we used higher-order DD with inter-pulse spacing $\tau < \tau_{surf} = 5$ μs at longer precession times. We focused on shallow spins as they are critical for nanoscale magnetometry. Figure 4 plots the coherence time, decoupling efficiency $\lambda$, and sensitivity versus number of pulses $N$ for NV m50 ($d = 9$ nm). The good agreement between experiment and numerical calculations using Eqs. 1 and 2 further supports the two-bath model. We observed a $T_{2,N=16}$ of 190(10) μs, which approaches the relatively short $T_1 = 300(10)$ μs. We note that $\lambda$ also increased with $N$. Importantly, sensitivity scales $\propto T_2^{-1/2}$ for spin detection schemes compatible with DD of the NV spin. A chief example is sensing a single proton spin via the oscillation of its transverse spin component at the Larmor frequency $v_p$, which is commonly in the kHz to few MHz range. In Fig. 4(c) we plot the expected sensitivity $\eta$ to a random phase magnetic field having correlation function $\langle \tilde{B}(t)\tilde{B}(t')\rangle = B_{rms}^2 \exp(-|t-t'|/T_c)\cos(2\pi v_p(t-t'))$ and phase correlation time $T_c = 1000$ μs, though the exact value of $T_c$ has little effect if $T_c > T_{2,N}$ [24]. $\eta$ is improved two-fold with just $N$ = 16 and approaches the value required to detect a single proton spin at the surface with SNR = 1 in 1 second. The long $T_1 >$ 1 ms measured for several doped NVs at even shallower depths ($d < 9$ nm) demonstrates the potential for increasing $T_2$ further.

We have presented a detailed study of decoherence of shallow NVs in a nitrogen delta-doped diamond film. The surface noise is well-modeled by a 2D electronic spin layer with sub-MHz dynamics, as evidenced by the depth dependence of coherence enhancement and total noise power probed by NVs at independently measured depths. We have shown that the decohering effects of fluctuating surface and bulk spins in nitrogen delta-doped diamond are mitigated via dynamical decoupling with appropriately chosen inter-pulse timing, which has significant impact for nano MRI and coherent spin coupling applications. The extracted $\sigma_{surf} = 0.04(2)$ spins/nm$^2$ is comparable to the densities found in experiments on metallic and insulating films [21]; this apparently universal phenomenon further emphasizes the need to identify the nature of these spins and the mechanism of the bath fluctuations. The scanning magnetic gradient method used here has recently facilitated high resolution NV based MRI of dark spins [29], making NVs an excellent sub-nm spectroscopic probe of this spin noise apparent in a variety of crystal surfaces.

Remaining questions about the diamond surface can be addressed using our method of shallow NV creation via growth combined with nanoscale depth imaging. Firstly, depth-calibrated studies of shallower NVs ($< 5$ nm) may reveal wide variations in $T_2$ from discrete surface spin effects or spin clustering [24]. Secondly, using delta-doping to form a dark nitrogen spin layer isolated from the diamond surface – $d > 60$ nm based on our findings – could provide a controlled test bed to study 2D spin bath effects on an NV outside the layer. Thirdly, under our present applied magnetic field we expect that NV coupling to electric and strain fields is of 2[nd] order [30], although experiments at $B_\parallel \approx 0$ could probe these effects near the surface. Lastly, we have presented a two-level dephasing model, but the incompletely understood $T_1$ processes between the $S = 1$ NV sublevels ultimately limit DD as a sensing protocol [31]. $T_1$ measurements of bulk [32] and shallow [33] NVs at lower temperatures suggest thermally activated relaxation rates of surface spins, and a future depth-calibrated study of both $T_1$ and $T_2$ at variable temperature could clarify the mechanism behind surface spin fluctuations or point to other sources of decoherence.

This work was supported by the DARPA QuASAR program and the AFOSR YIP. B.A.M. is supported through a fellowship from the Department of Defense (NDSEG). The authors thank V. V. Dobrovitski, K. Lee, C. McLellan, and L. Pascal for helpful discussions.

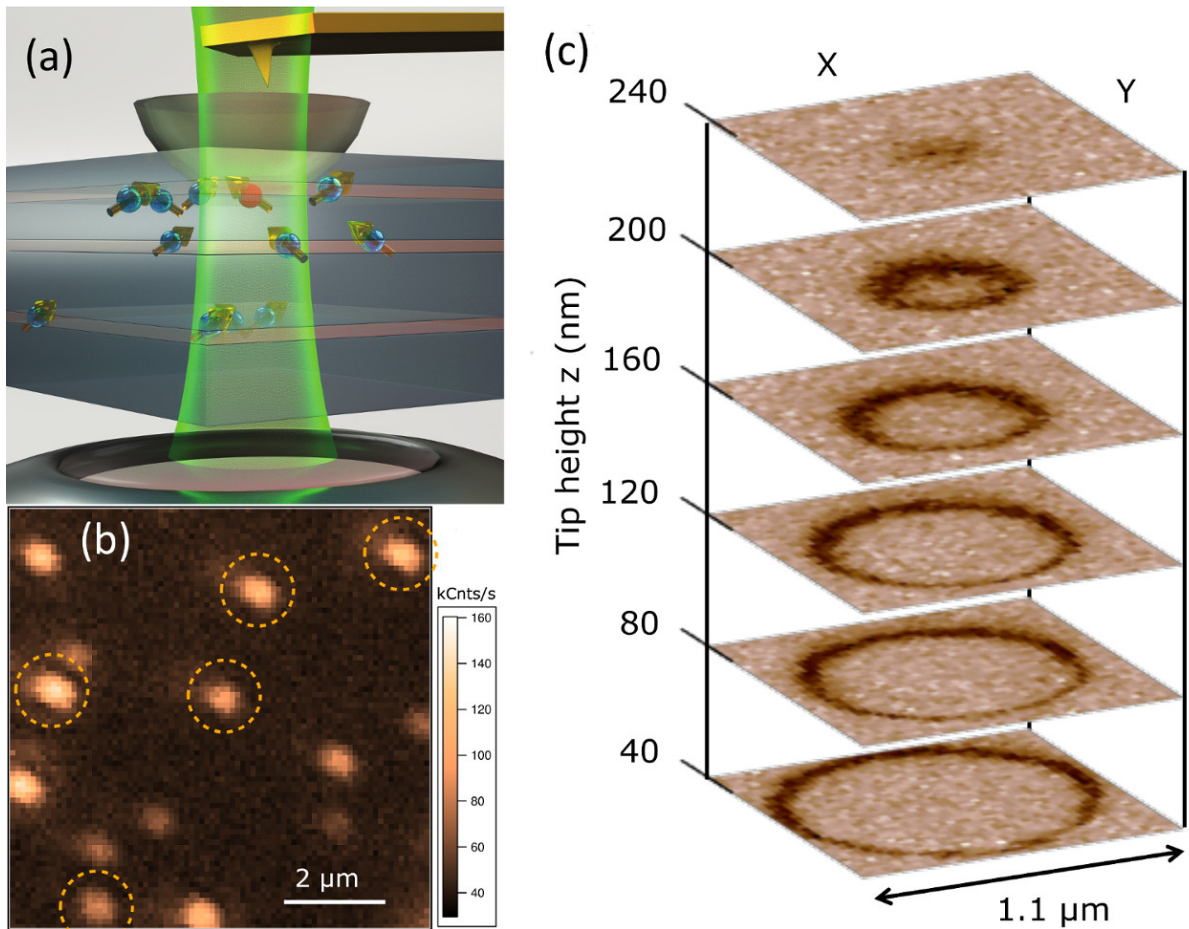

Figure 1: (a) Schematic of a CVD-grown diamond film with three nitrogen delta-doped layers (orange) that contain nitrogen-vacancy (NV) spins at nanoscale separations. A 532 nm laser is focused onto NVs via an inverted confocal microscope giving a diffraction-limited depth of field. To achieve nanoscale depth discrimination between NVs, a scanning magnetic tip and microwave field form a resonance slice. Colored red is an NV that intersects this slice. (b) Confocal image showing individual NVs. (c) Optically detected ESR images recorded as a function of the magnetic tip position over a single NV. Dark rings mark reduced fluorescence when the ESR slice crosses the NV.

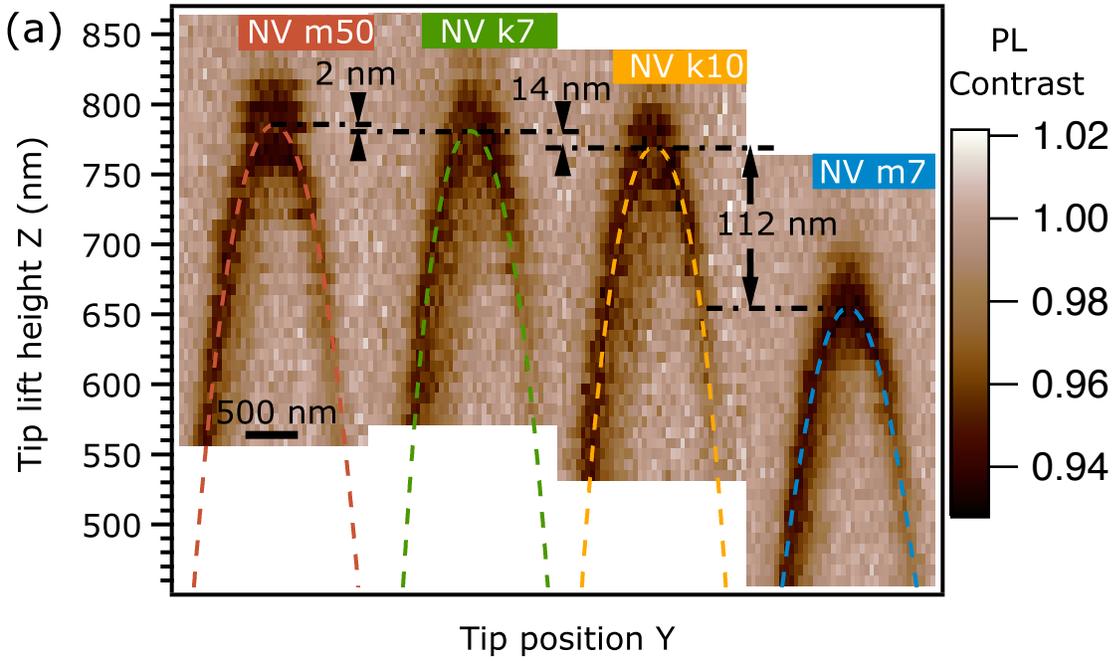
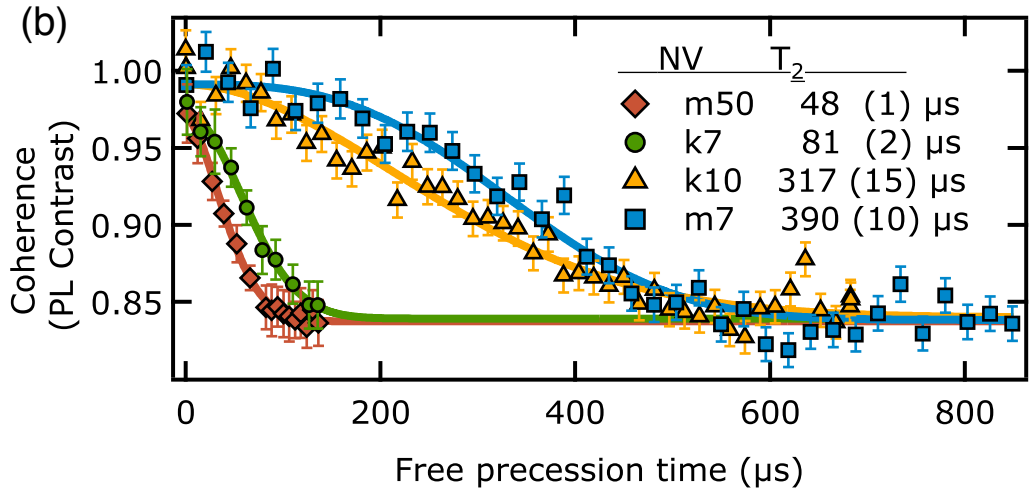

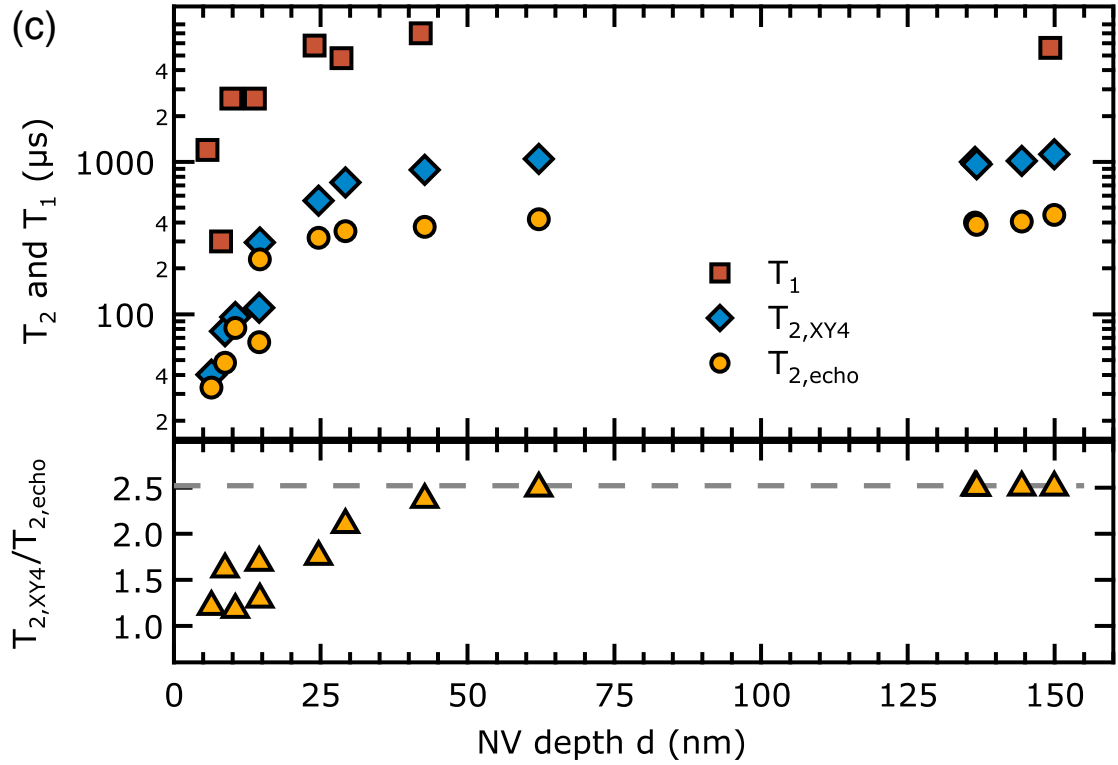

Figure 2: (a) ESR slice photoluminescence (PL) contrast images in a lateral-height plane $(YZ)$ measured for four NVs of identical orientation. Relative depths were extracted via image registration [24], and dashed curves are polynomial fits as guides to the eye. (b) Hahn echo coherence data (markers) with shot noise-limited error bars and fits (lines) for the four NVs in (a). (c) Coherence times ($T_{2,\text{echo}}$ and $T_{2,\text{XY4}}$) and relaxation times ($T_1$) versus NV depth, showing strong suppression of coherence near the surface. The lower panel shows $T_{2,\text{XY4}}/T_{2,\text{echo}} = N^\lambda$ is reduced with decreased depth, indicating that dynamical decoupling with $N = 4$ pulses is less efficient for shallower NVs. The dashed line $(\lambda = 2/3)$ is expected for NV dephasing by a slow bulk nitrogen spin bath.

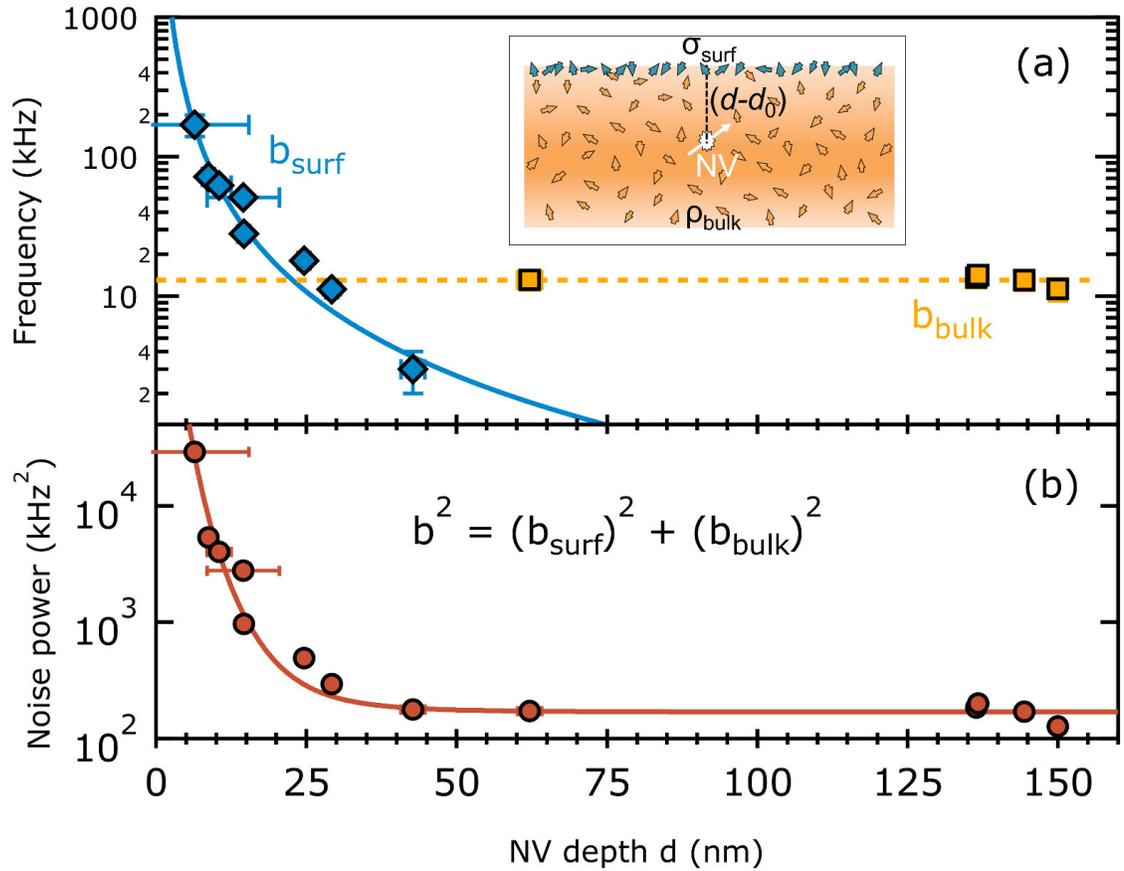

Figure 3: (a) Depth dependence of NVs' coupling frequency to the surface (blue diamonds) and bulk (orange squares) noise baths as extracted by fitting coherence decay data to a two-bath dephasing model (see inset schematic). The solid blue curve is a fit to a 2D electronic spin bath model. The fit gives absolute NV depths (shallowest 6 nm), nitrogen spin density $\rho_{bulk} = 8.6 \times 10^{15}$ cm$^{-3}$, and surface spin density $\sigma_{surf} = 0.04$ nm$^{-2}$. b) Integrated noise power of NV-bath coupling frequency. The solid line is a fit to the two-bath model.

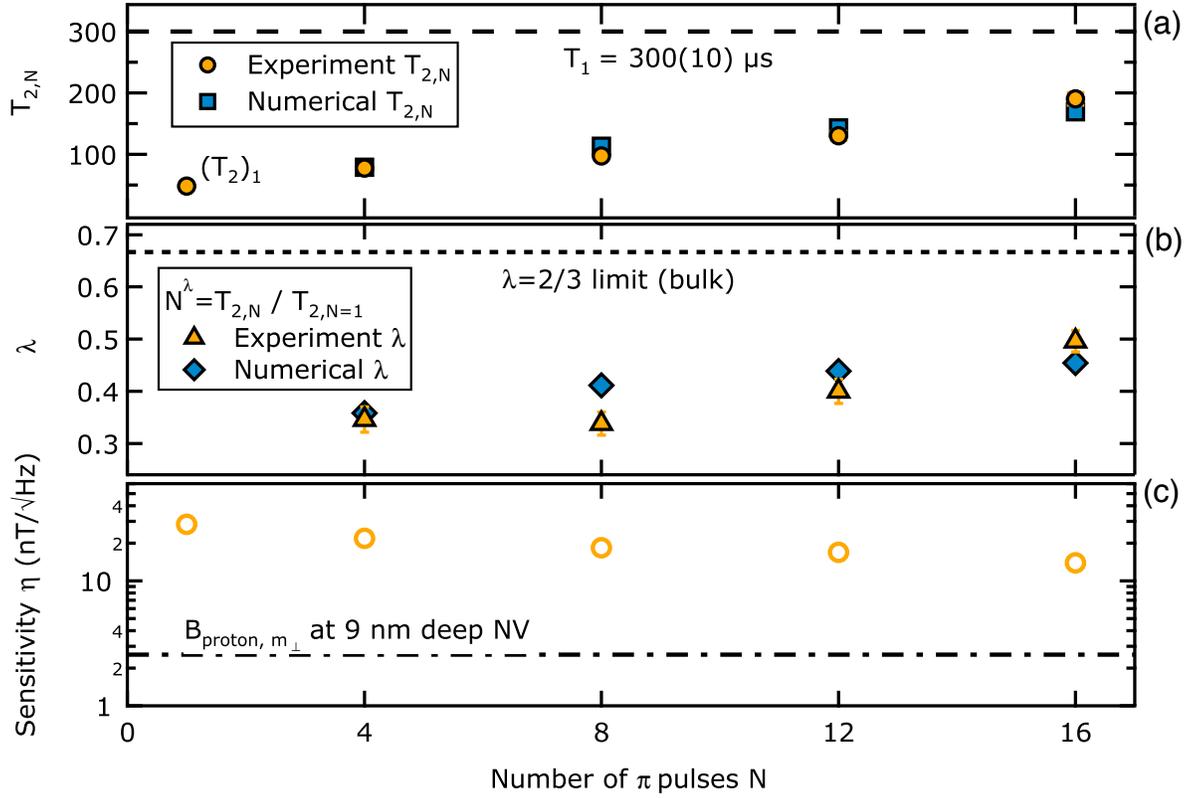

Figure 4: Effects of increased number of pulses $N$ in XY4 dynamical decoupling of a 9 nm deep NV. (a) Measured $T_2$ (orange circles) and numerical calculations based on the dephasing due to surface and bulk spins (blue squares). (b) The measured decoupling efficiency parameter $\lambda$ (orange triangles) also improves as $N$ increases, confirming that our deduced $\tau_{surf} \sim 5\ \mu s$ is accessible to practical decoupling inter-pulse times. (c) Magnetic sensitivity $\eta$ of the NV approaches a few $\mathrm{nT}/\sqrt{\mathrm{Hz}}$; the dash-dot line marks the $\eta$ required to detect a single proton at the surface in 1 second with $\mathrm{SNR}=1$.